\documentclass[traditabstract, letter]{aa} 
\usepackage{graphicx}
\usepackage{txfonts}
\begin{document}

   \title{Discovery of water vapour in the carbon star V Cygni from observations with  Herschel/HIFI\thanks{Herschel is an ESA space observatory with science instruments provided
by European-led Principal Investigator consortia and with important participation from NASA.}}

   \author{
D.~A. Neufeld\inst{1} \and
E.~Gonz\'alez-Alfonso\inst{2} \and
G.~Melnick\inst{3} \and 
M.~Pu{\l}ecka\inst{4} \and
M.~Schmidt\inst{4} \and
R.~Szczerba\inst{4} \and
V.~Bujarrabal\inst{5}\and 
J.~Alcolea\inst{6}\and
J.~Cernicharo\inst{7}\and
L.~Decin\inst{8,9}\and
C.~Dominik \inst{9,15}\and
K.~Justtanont\inst{10}\and
A.~de~Koter\inst{9,16}\and
A.~P.~Marston\inst{11}\and
K.~Menten\inst{12}\and
H.~Olofsson\inst{10,13}\and
P.~Planesas\inst{5,14}\and
F.~L.~Sch\"oier\inst{10}\and
D.~Teyssier\inst{11}\and
L.~B.~F.~M.~Waters\inst{9,8}\and
K.~Edwards\inst{17}\and                   
C.~McCoey\inst{17}\and 
R.~Shipman\inst{18}\and                
W.~Jellema\inst{18}\and     
T.~de~Graauw\inst{19}\and
V.~Ossenkopf \inst{20}\and
R.~Schieder \inst{20}\and
S.~Philipp \inst{12}}


   \institute{ The Johns Hopkins University, 3400 North Charles St, Baltimore, MD 21218, USA \\ 
\email{neufeld@pha.jhu.edu}  
\and Departamento de F\'{\i}sica, Universidad de Alcal\'a de Henares, Campus Universitario, E-28871 Alcal\'a de Henares, Madrid, Spain 
\and Harvard-Smithsonian Center for Astrophysics, Cambridge, MA 02138, USA 
\and N. Copernicus Astronomical Center, Rabia{\'n}ska 8, 87-100 Toru{\'n}, Poland 
\and Observatorio Astron\'omico Nacional (IGN), Ap 112, E--28803 Alcal\'a de Henares, Spain 
\and Observatorio Astron\'omico Nacional (IGN), Alfonso XII N$^{\circ}$3,E--28014 Madrid, Spain 
\and CAB, INTA-CSIC, Ctra de Torrej\'on a Ajalvir, km 4,E--28850 Torrej\'on de Ardoz, Madrid, Spain 
\and Instituut voor Sterrenkunde,Katholieke Universiteit Leuven, Celestijnenlaan 200D, 3001 Leuven, Belgium 
\and Sterrenkundig Instituut Anton Pannekoek, University of Amsterdam, Science Park 904, NL-1098 Amsterdam, The Netherlands 
\and Onsala Space Observatory, Dept. of Radio and Space Science, Chalmers  University of Technology, SE--43992 Onsala, Sweden 
\and European Space Astronomy Centre, ESA, P.O. Box 78, E--28691 Villanueva de la Ca\~nada, Madrid, Spain 
\and Max-Planck-Institut f{\"u}r Radioastronomie, Auf dem H{\"u}gel 69, D-53121 Bonn, Germany 
\and Department of Astronomy, AlbaNova University Center, Stockholm University, SE--10691 Stockholm, Sweden 
\and Joint ALMA Observatory, El Golf 40, Las Condes, Santiago, Chile 
\and Department of Astrophysics/IMAPP, Radboud University Nijmegen,    Nijmegen, The Netherlands 
\and Astronomical Institute, Utrecht University, Princetonplein 5, 3584 CC Utrecht, The Netherlands 
\and University of Waterloo, and University of Western Ontario, Canada 
\and SRON Netherlands Institute for Space Research, Landleven 12, 9747 AD Groningen, Netherlands 
\and Atacama Large Millimeter/Submillimeter Array, Joint ALMA Office, Santiago, Chile 
\and KOSMA, I. Physik. Institut, Universit\"at zu K\"oln, Z\"ulpicher Str. 77, D 50937 K\"oln, Germany} 

  \abstract
{We report the discovery of water vapour toward the carbon star V Cygni.  We have used Herschel's HIFI instrument, in dual beam switch mode, to observe the $1_{11} - 0_{00}$ para-water transition at 1113.3430~GHz in the upper sideband of the Band 4b receiver.  The observed spectral line profile is nearly parabolic, but with a slight asymmetry associated with blueshifted absorption, and the integrated antenna temperature is $1.69 \pm 0.17$ K~km~s$^{-1}$.  This detection of thermal water vapour emission, carried out as part of a small survey of water in carbon-rich stars, is only the second such detection toward a carbon-rich AGB star, the first having been obtained by the 
{\it Submillimeter Wave Astronomy Satellite} toward IRC+10216.  For an assumed ortho-to-para ratio of 3 for water, the observed line intensity implies a water outflow rate $\sim 3 - 6 \times 10^{-5}$ Earth masses per year and a water abundance relative to H$_2$ of $\sim 2 - 5 \times 10^{-6}$.  This value is a factor of at least $10^4$ larger than the expected photospheric abundance in a carbon-rich environment,  and -- as in IRC+10216 --  raises the intriguing {\it possibility} that the observed water is produced by the vapourisation of orbiting comets or dwarf planets.  {  However,} observations of the single line observed to date do not permit us to place strong constraints upon the spatial distribution {  or origin} of the observed water, but future observations of additional transitions will allow us to determine the inner radius of the H$_2$O-emitting zone, and the H$_2$O ortho-to-para ratio, and thereby to place important constraints upon the origin of the observed water emission.}  

               
   \titlerunning{Water vapour in V Cygni}
	\authorrunning{Neufeld et al.}
   \maketitle
%

\section{Introduction}

The carbon-to-oxygen ratio is the critical determinant of the photospheric chemistry in evolved stars; the photospheres of oxygen rich-stars, with C/O ratios $<$ 1, are dominated by CO and H$_2$O, while those of carbon-rich stars are dominated by CO, HCN, and C$_2$H$_2$ and contain very little H$_2$O.
Despite this sharp dichotomy in photospheric composition, water vapour has
previously been discovered, using the Submillimeter Wave Astronomy Satellite
(SWAS), in the circumstellar outflow of the extreme carbon star IRC+10216
(Melnick et al.\ 2001), for which the C/O ratio is $\sim 1.4$.
The observation of water vapour in that source, with an inferred abundance
relative to H$_2$ of $\sim 10^{-7}$ (Ag\'undez \& Cernicharo 2006; Gonz\'alez-
Alfonso, Neufeld \& Melnick 2007), has led to the suggestion of several possible
origins for the water vapour, including (1) the vapourisation of icy objects (comets or dwarf planets) in orbit around the
star (Ford \& Neufeld 2001);
(2) Fischer-Tropsch catalysis (Willacy 2004); (3) photochemistry within an outer, photodissociated shell (Ag\'undez \& Cernicharo 2006);
(4) photochemistry within a clumpy outflow (Decin et al.\ 2010).  {  In both carbon- and oxygen-rich stars,  Cherchneff (2006) and Decin et al.\ (2008) have argued for the importance of shocks in setting non-equilibrium abundances of the molecules CO, HCN, CS, and SiO within the inner envelope, although existing models do not appear to predict a persistent enhancement of H$_2$O in stars with C/O $> 1$.}

Using the HIFI instrument (de~Graauw et al.\ 2010) on Herschel (Pilbratt et al.\ 2010), it is now possible to search with unprecedented sensitivity for water vapour in carbon-rich stars. As part of the HIFISTARS Guaranteed Time Key program, we will
carry out a survey for water vapour in eight carbon-rich stars at distances
greater than that ($\sim 170$~pc) of IRC+10216. Here, we report the detection of
water vapour in the very first carbon star targeted in this survey: V Cygni, a Mira variable of spectral type C6 (Wallerstein \& Knapp 1998), with an apparent V magnitude varying from 13.9 to 7.7 mag, a period of 421 days, and a distance of $\sim$ 400 pc (Bieging \& Wilson 2001).   

\section{Observations and results}

We observed the lowest transition of para-water, the $1_{11} - 0_{00}$ transition at 1113.3430~GHz line, in the upper sideband of the Band 4b HIFI receiver.  This observation, of duration 1167~s including overheads, was carried out on 2010 March 5, using the dual beam switch (DBS) mode and the Wide Band Spectrometer (WBS).  The WBS has a spectral resolution of 1.1~MHz, corresponding to a velocity resolution of $0.30\, \rm km\,s^{-1}$ at the frequency of the $1_{11} - 0_{00}$ transition. The telescope beam, of half-power-beam-width (HPBW) $\sim 20^{\prime\prime}$, was centered on V Cygni at coordinates 
$\rm \alpha=20h\,41m\,18.3s, \delta = 48^0\, 08^\prime \,29^{\prime\prime} (J2000)$.  
The reference positions for this observation were located at offsets of $3^{\prime}$ on either side of the source.

The data were processed using the standard HIFI pipeline to Level 2, providing fully calibrated spectra of the source.  The Level 2 data were analysed further using the Herschel Interactive Processing Environment (HIPE; Ott 2010), version 2.4, along with ancillary IDL routines that we have developed.  Having found the signals measured in the two orthogonal polarizations to be in excellent agreement, we combined them to obtain an average spectrum. 
Figure 1 shows the WBS spectrum of para-H$_2$O $1_{11}-0_{00}$ obtained toward V Cygni, with the frequency scale expressed as Doppler velocities relative to the Local Standard of Rest (LSR) and the intensity scale expressed as antenna temperature.  {   A zeroth-order baseline has been subtracted}.
The vertical dashed line indicates the LSR velocity of the source, as determined by Bieging \& Wilson (2001, hereafter BW01) from observations of the CO $J=2-1$ line.  The integrated antenna 
temperature is found to be $1.69 \pm 0.17$~K~km~s$^{-1}$.  {  Our identification of the observed feature with the para-H$_2$O $1_{11}-0_{00}$ transition is supported by the absence of any plausible alternative candidate in the JPL (Pickett et al.\ 1998) or CDMS (M{\"u}ller et 
al.\ 2001) spectral line catalogues.}

\begin{figure}
\includegraphics[width=9 cm]{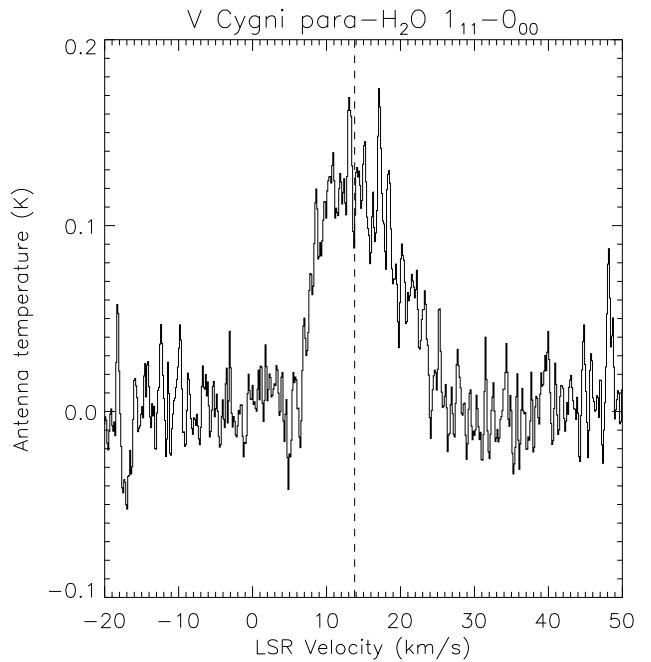}
\caption{Continuum-subtracted spectrum of para-H$_2$O $1_{11}-0_{00}$ obtained toward V Cygni.  The vertical dashed line indicates the LSR velocity of the source, as determined by BW01 from observations of the CO $J=2-1$ line.}

\end{figure}

\begin{table}
\caption{Assumed parameters for V Cygni}
\begin{tabular}{l l}
\hline\hline
Distance: 400 pc \\
Stellar luminosity: 8500~$L_{\odot}$ \\
Effective temperature: 2500 K \\
Stellar radius: $3.42 \times 10^{13}$~cm \\
Systemic LSR velocity: $+13.8 \rm \, km \,s^{-1}$ \\
Terminal outflow velocity: $11.8 \rm \,  km\, s^{-1}$ \\
\hline
\end{tabular}
\end{table}

\section{Derivation of the water outflow rate and abundance}

In modeling the water emission from V Cygni, we have used the methods described by Gonz\'alez-Alfonso, Neufeld \& Melnick (2007, hereafter GNM).  To determine the H$_2$O rotational populations and the resultant H$_2$O emission spectrum, we included the effects of radiative pumping -- through the $\nu_2=1$ and $\nu_3=1$ vibrational states -- and of collisional excitation by H$_2$, together with a treatment of radiative transfer based upon that discussed by Gonz\'alez-Alfonso \& Cernicharo (1997).

The assumed parameters for the source are summarized in Table 1 and discussed below.  A variety of distance estimates for V Cygni have appeared in the literature.  BW01 obtained values of 406 pc and 456 pc respectively by using the $P-K$ (period -- absolute K magnitude) and $P-L$ (Period -- bolometric luminosity) relationships presented by Groenewegen \& Whitelock (1996; hereafter GW96).  However, our own fit to the spectral energy distribution (SED), shown in Figure 2, implies that a distance of 342~pc is required to fit the GW96 $P-L$ relationship.  Here, we used a dust radiative transfer model to fit a combination of flux measurements from {\it Tycho}, {\it Hipparcos}, {\it MSX}, 2MASS, {\it ISO}, {\it IRAS}, and JCMT.  This model, full details of which will be presented in a future publication (Schmidt et al.\ 2010), assumes a power-law distribution of grain radii with index $-3.5$ between 0.005 and 0.25 $\mu$m, a maximum dust temperature of 1050~K, an inner radius for the dust shell of $2 \times 10^{14}$~cm, and a total optical depth $A_V = 9.45$~mag. A smaller distance estimate (370~pc) than those of BW01  was also obtained by Sch\"oier \& Olofsson (2000).  All of the values mentioned above are consistent with the {\it Hipparcos}-measured parallax of $3.69 \pm 1.77$ mas, which implies a $1\sigma$ distance range of 180 -- 520 pc.  Accordingly, we adopt a value of 400~pc for the estimated distance of V Cygni, with a likely uncertainty $~\sim 20\%$.  

\begin{figure}
\includegraphics[width=9 cm]{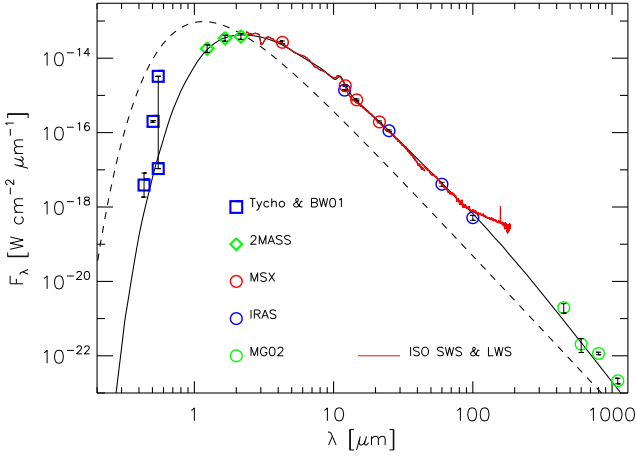}
\caption{Spectral energy distribution of V Cygni.  A dust radiative transfer model (see text) has been used to fit the fluxes measured by {\it Tycho},  {\it MSX}, 2MASS, {\it ISO}, {\it IRAS}, and JCMT.  The dashed curve shows the unattenuated photospheric emission, while the solid curve shows the total emission from the attenuated photosphere and surrounding dust.}
\end{figure}

This estimate of the distance then requires a bolometric luminosity of $8500\,L_\odot$ to match the observed SED shown in Figure 2, and a stellar radius of $3.42 \times 10^{13} \, \rm cm,$ given an assumed effective temperature of 2500~K.  The values given in Table 1 for the systemic velocity of the source and the terminal velocity of the outflow were based upon {\it Herschel}/HIFI observations of the CO $J=6-5$ transition at 691.473~GHz (Schmidt et al.\ 2010) and are in excellent agreement with previous determinations by BW01.  To obtain an estimate of the total mass-loss rate in the outflow, we have modeled the fluxes observed by {\it Herschel}/HIFI for the CO $J=6-5$, $J=10-9$ and $J=16-15$ transitions.  Full details of the observations and modeling of CO will be given by Schmidt et al.\ (2010).  Our best estimate of the gas mass-loss rate in the inner envelope is $4.6 \times 10^{-6} M_\odot\, \rm yr^{-1},$ a factor $\sim 3$ larger than that obtained by Sch\"oier \& Olofsson (2000) from a fit to the CO $J=2-1$ transition, and the derived CO/H$_2$ ratio is 10$^{-3}$.  Indeed, we find that a constant mass-loss rate model that fits the CO $J=6-5$, $J=10-9$ and $J=16-15$ transitions substantially overpredicts the flux in the CO $J=2-1$ transition. 
This discrepancy suggests some variability in the mass-loss rate, with a larger value applying to the inner envelope where the higher-lying transitions of CO originate.  We obtained a satisfactory fit to both the CO rotational line fluxes and the continuum spectrum by assuming a gas and dust density that decreases as radius$^{-2.15}$, instead of the radius$^{-2}$ density profile expected for an envelope with a constant mass loss rate and outflow velocity.  The gas-to-dust mass ratio in this model is 510.  

Fortunately, given the significant uncertainties in many of the assumed parameters listed in Table 1, the derived water outflow rate is not strongly dependent upon any of them.  As discussed in GNM, unless the mass-loss rate is extremely large, the excitation of water is dominated by radiative pumping via the 6$\,\mu$m $\nu_2$ band.  Thus, for a given water outflow rate, the observed water line fluxes scale linearly with the observed 6$\,\mu$m continuum flux.  Since the latter is an observed (rather than a derived) quantity, our estimate of the water outflow rate is largely independent of the distance or total outflow rate assumed for the source.

\begin{figure}
\includegraphics[width=9 cm]{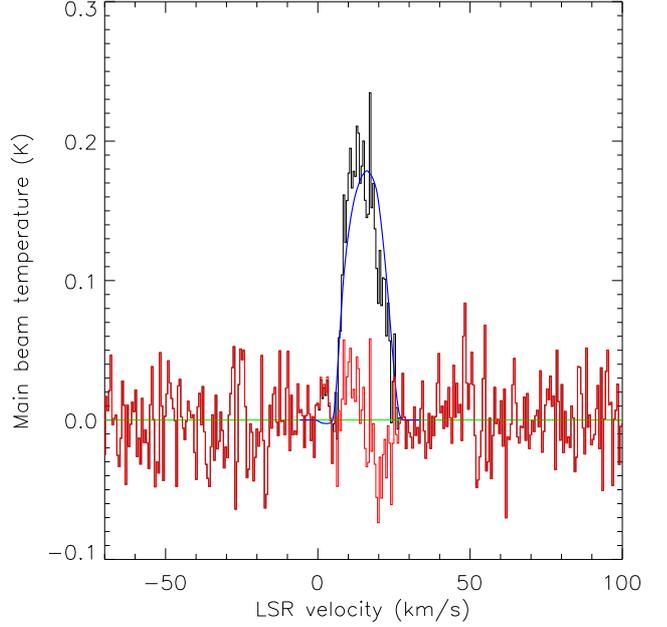}
\caption{Continuum-subtracted and smoothed spectrum of the $1_{11}- 0_{00}$ transition of para-water at 1113.3430 GHz (black), compared with the the best fit from Model B (blue).  The residuals are shown in red.  Here, the vertical axis is the main beam brigthness temperature, obtained by dividing the antenna temperature by an assumed main beam efficiency of 0.67}
\end{figure}

In modeling the water line strength and profile observed toward V Cygni, we have investigated two models for the spatial distribution of the observed water.  In Model A, we assume that water is present at radii as small as $4.5 \times 10^{14}$~cm, while in Model B, we adopt an inner radius $R_{\rm in} = 2 \times 10^{15} \, \rm cm$ at which the water is injected into the outflow.  In both cases, we assume an outer radius of  $R_{\rm out} = 1 \times 10^{16} \, \rm cm$, the estimated photodissociation radius for water.\footnote{The mean lifetime against photodissociation for a water molecule in the unshielded mean Galactic ultraviolet radiation field is $1.2 \times 10^9\,\rm s$ (Roberge et al.\ 1991), corresponding to a distance of $1.4 \times 10^{15} \rm \,cm$ at the terminal outflow velocity of the V Cygni outflow.  Shielding by dust in the outflow increases this distance by a factor of several, leading to photodissociation radii ranging from ~$\sim 5 - 10 \times 10^{15}\, \rm cm^{-2}$ for radiation fields within a factor of two of the mean Galactic value.  We adopt the upper end of this range, in order to obtain a conservative estimate of the water outflow rate.  To account for the observed water line strength, smaller values of $R_{\rm out}$ would require larger values of the water abundance.}

The distribution assumed in Model B is expected if the vapourisation of icy objects is the origin of the observed water vapour, since all such objects at smaller distances from the star will have been vapourised already (Ford \& Neufeld 2001, GNM).  In both models, we adopted the profile given by Goldreich \& Scoville (1976) for the outflow velocity as a function of radius.  For Model B, this velocity profile was found to give a better fit to the observed line shape than an alternative velocity profile, based upon the work of Skinner et al.\ (1999), which entails a more rapid acceleration of the outflowing material.
A clear asymmetry in the line profile, apparently associated with blueshifted absorption, suggests a significant turbulent velocity width, of order $1.3 \, \rm km \, s \, ^{-1}$.  {\it Equally good fits can be obtained for models A and B, implying that our current observations (of a single line) cannot be used to provide useful constraints on the spatial distribution {  or origin} of the water vapour in V Cygni.}  The best-fit parameters are given in Table 2, for each of the two models, and the fit for Model B is shown in Figure 3.  For an assumed water ortho-to-para ratio of 3, the required water outflow rates are $3 \times 10^{-5}$ and $6 \times 10^{-5}$ Earth masses per year for models A and B respectively. 

\begin{table}
\caption{Derived parameters for V Cygni}
\begin{tabular}{l l}
\hline\hline
{Model A}: Water inner radius = $4.5 \times 10^{14} \, \rm cm$ \\
\\
Para-water abundance relative to H$_2$: $5.3 \times 10^{-7}$ \\
Total water abundance$^a$ relative to H$_2$: $2.1 \times 10^{-6}$ \\
Total water outflow rate: $3 \times 10^{-5}$ Earth mass / yr \\
\\
{Model B}: Water inner radius = $2 \times 10^{15} \, \rm cm$ \\
\\
Para-water abundance relative to H$_2$: $1.0 \times 10^{-6}$ \\
Total water abundance$^a$ relative to H$_2$: $4.2 \times 10^{-6}$ \\
Total water outflow rate: $6 \times 10^{-5}$ Earth mass / yr \\
\hline
\hline
$^a$for an assumed ortho-to-para ratio of 3 \\
\end{tabular}
\end{table}

\section{Discussion}

Our detection of thermal water vapour from V Cygni is only the second such detection from the circumstellar envelope of a carbon-rich AGB star, the first having been obtained toward IRC+10216.
The water outflow rate in V Cygni, $3 - 6 \times 10^{-5}$ Earth masses per year, is 4 -- 8 times that in IRC+10216, even though the total mass outflow rate of $4.6 \times 10^{-6} M_\odot\, \rm yr^{-1}$ is smaller by a factor $\sim 8$.  Thus the water line luminosity in V Cygni implies a water abundance $\sim 30 - 60$ times as large as that inferred for IRC+10216.  The large H$_2$O abundance $\sim 3 - 6 \times 10^{-6}$ that we infer is at least a factor of 10$^4$ larger than that predicted in models for the photospheres of C-rich stars {  in thermochemical equilibrium (e.g.\ Cherchneff 2006)}.  As in IRC+10216, the inferred abundance raises the intriguing {\it possibility} that the water is produced by the vapourisation of orbiting comets or dwarf planets. In this picture (Ford \& Neufeld 2001), a collection of orbiting icy objects in a Kuiper-belt analogue are vapourised as the star ascends the asymptotic giant branch and its luminosity gradually increases; to match the observed water outflow rate in V Cygni, this scenario would require a reservoir of $\sim 10$ Earth masses of water ice available for release during the AGB phase.

Observations of the single water line observed to date, {  however,} do not permit us to place strong constraints upon the spatial distribution {  or origin} of the observed water, but future observations of additional transitions will allow us to determine the inner radius, $R_{\rm in}$, of the H$_2$O-emitting zone, as well as the H$_2$O ortho-to-para ratio.  The radius, $R_{\rm in}$, is a critical diagnostic, because the comet vapourisation model predicts a value $\sim 2 \times 10^{15} \rm \, cm$ (Model B), all orbiting icy objects within that radius having already been vapourised.  In Table 3, we present predictions for the strengths of various emission lines lying within the spectral range accessible to the HIFI, SPIRE and PACS instruments on {\it Herschel}.  In the HIFI (and SPIRE) spectral range (upper section of Table 3), results are given as integrated antenna temperatures; in the PACS spectral range (lower section), the values given are fluxes, in units 10$^{-16}$~W~m$^{-2}$.  The predictions presented here are for an assumed ortho-to-para ratio of 3.  Several lines, such as the 916.171~GHz $4_{22}-3_{31}$ transition observable by HIFI and the $6_{16}-5_{05}$ 82.031~$\mu$m transition observable by PACS, show a strong dependence upon $R_{\rm in}$  and should provide an excellent discriminant between different models.

\begin{table}
\caption{Predicted line strengths for V Cygni}
\begin{tabular}{c r r c c}
\hline\hline
\\
Transition & Frequency & Wavelength & \multicolumn{2}{c}{Line strength}\\
           & (GHz)     & ($\mu$m)   & \multicolumn{2}{c}{$\int T_{MB}\, dv$ in K~$\rm km\,s^{-1}$ $\,^a$} \\
           &           &            & Model A & Model B\\

\hline
$1_{11}-0_{00}$ & 1113.343 &   269.272 &   2.53$^a$ &   2.53$^b$\\
$2_{02}-1_{11}$ &    987.927 &   303.456 & 2.46  &   2.79\\
$2_{11}-2_{02}$ &    752.033 &   398.643 & 1.18 &    1.17\\
$2_{20}-2_{11}$ &  1228.789 &   243.974 &  0.56 &   0.56\\
$4_{22}-3_{31}$ &    916.171 &   327.223 & 0.47 &   0.02\\
$1_{10}-1_{01}$ &  556.936 &   538.289 &   1.16 &   1.06\\
$2_{12}-1_{01}$ &  1669.905 &   179.527 &  6.40 &   5.90\\
$2_{21}-2_{12}$ & 1661.008 &   180.488 &   1.92 &   1.79\\
$3_{03}-2_{12}$ & 1716.770 &   174.626 &   5.43 &   4.93\\
$3_{12}-2_{21}$ & 1153.126 &   259.982 &   1.37 &   0.85\\
$3_{12}-3_{03}$ & 1097.365 &   273.193 &   1.67 &   1.49\\
$3_{21}-3_{12}$ & 1162.911 &   257.795 &   0.65 &   0.59\\
$4_{32}-5_{05}$ &  1713.882 &   174.920 &  0.64 &   0.03\\
\hline
\\
Transition & Frequency & Wavelength & \multicolumn{2}{c}{Line strength}\\
           & (GHz)     & ($\mu$m)   & \multicolumn{2}{c}{($10^{-16}\,\rm W\,m^{-2}$)} \\
           &           &            & Model A & Model B\\

\hline
$2_{11}-2_{02}$ & 2968.748 &   100.983 &      1.27 &      0.82\\
$3_{31}-2_{20}$ & 4468.572 &    67.089 &      1.16 &      0.42\\
$2_{21}-1_{10}$ & 2773.976 &   108.073 &      2.20 &      1.67\\
$3_{21}-2_{12}$ & 3977.045 &    75.381 &      2.61 &      1.11\\
$4_{14}-3_{03}$ & 2640.473 &   113.537 &      1.70 &      0.70\\
$3_{30}-2_{21}$ & 4512.385 &    66.438 &      2.51 &      1.12\\
$4_{32}-3_{12}$ & 3807.256 &    78.742 &      1.72 &      0.23\\
$5_{05}-4_{14}$ & 3013.199 &    99.493 &      1.23 &      0.22\\
$4_{32}-3_{21}$ & 5107.284 &    58.699 &      1.73 &      0.35\\
$6_{16}-5_{05}$ & 3654.602 &    82.031 &      1.20 &      0.02\\

\hline
\hline
\\
\end{tabular}
\noindent $\,$ $^a$integrated main beam temperature

\noindent $^b$observed line strength
\end{table}

\appendix

\section{Acknowledgements}
HIFI has been designed and built by a consortium of institutes and university departments from across
Europe, Canada and the United States under the leadership of SRON Netherlands Institute for Space
Research, Groningen, The Netherlands and with major contributions from Germany, France and the US.
Consortium members are: Canada: CSA, U.~Waterloo; France: CESR, LAB, LERMA, IRAM; Germany:
KOSMA, MPIfR, MPS; Ireland, NUI Maynooth; Italy: ASI, IFSI-INAF, Osservatorio Astrofisico di Arcetri-
INAF; Netherlands: SRON, TUD; Poland: CAMK, CBK; Spain: Observatorio Astron\'omico Nacional (IGN),
Centro de Astrobiolog\'a (CSIC-INTA). Sweden: Chalmers University of Technology - MC2, RSS \& GARD;
Onsala Space Observatory; Swedish National Space Board, Stockholm University - Stockholm Observatory;
Switzerland: ETH Zurich, FHNW; USA: Caltech, JPL, NHSC.

This research was performed, in part, through a JPL contract funded by the National Aeronautics and Space Administration.
R.Sz. and M.Sch. acknowledge support from grant N 203 393334 from Polish MNiSW.  E.G-A  is a Research Associate at the Harvard-Smithsonian 
Center for Astrophysics.  This work has been partially supported by the Spanish MICINN, program CONSOLIDER INGENIO 2010, grant ASTROMOL (CSD2009-00038).


\begin{thebibliography}{}

\bibitem[Ag{\'u}ndez 
\& Cernicharo(2006)]{2006ApJ...650..374A} Ag{\'u}ndez, M., \& Cernicharo, J.\ 2006, \apj, 650, 374 

\bibitem[Bieging 
\& Wilson(2001)]{2001AJ....122..979B} Bieging, J.~H., \& Wilson, C.~D.\ 2001, \aj, 122, 979 

\bibitem[Cherchneff(2006)]{2006A&A...456.1001C} Cherchneff, I.\ 2006, \aap, 456, 1001 

\bibitem[Decin et 
al.(2008)]{2008A&A...480..431D} Decin, L., Cherchneff, I., Hony, S., Dehaes, S., De Breuck, C., \& Menten, K.~M.\ 2008, \aap, 480, 431 

\bibitem[De]{De} Decin, L., Ag\'undez, M., Barlow, M. J., et al.\ 2010, \nat, in press

\bibitem[dG]{dG} De Graauw, T.\ et al.\, \aap, in press.

\bibitem[Ford 
\& Neufeld(2001)]{2001ApJ...557L.113S} Ford, K.~E., \& Neufeld, D.~A.\ 2001, \apjl, 557, L113 

\bibitem[Gonzalez-Alfonso 
\& Cernicharo(1997)]{1997A&A...322..938G} Gonz\'alez-Alfonso, E., \& Cernicharo, J.\ 1997, \aap, 322, 938 

\bibitem[Gonz{\'a}lez-Alfonso et al.(2007)]{2007ApJ...669..412G} 
Gonz{\'a}lez-Alfonso, E., Neufeld, D.~A., 
\& Melnick, G.~J.\ 2007, \apj, 669, 412 

\bibitem[Goldreich 
\& Scoville(1976)]{1976ApJ...205..144G} Goldreich, P., \& Scoville, N.\ 1976, \apj, 205, 144 

\bibitem[Groenewegen 
\& Whitelock(1996)]{1996MNRAS.281.1347G} Groenewegen, M.~A.~T., \& Whitelock, P.~A.\ 1996, \mnras, 281, 1347 

\bibitem[Melnick et al.(2001)]{2001Natur.412..160M} Melnick, G.~J., 
Neufeld, D.~A., Ford, K.~E.~S., Hollenbach, D.~J., 
\& Ashby, M.~L.~N.\ 2001, \nat, 412, 160

\bibitem[M{\"u}ller et 
al.(2001)]{2001A&A...370L..49M} M{\"u}ller, H.~S.~P., Thorwirth, S., Roth, D.~A., \& Winnewisser, G.\ 2001, \aap, 370, L49 

\bibitem[O]{O} Ott, S. 2010, in Astronomical Data Analysis Software and Systems XIX, ed. Y.
Mizumoto, K.-I. Morita, \& M. Ohishi, ASP Conf. Series

\bibitem[Pickett et al.(1998)]{1998JQSRT..60..883P} Pickett, H.~M., 
Poynter, I.~R.~L., Cohen, E.~A., Delitsky, M.~L., Pearson, J.~C., 
\& M{\"u}ller, H.~S.~P.\ 1998, Journal of Quantitative Spectroscopy and Radiative Transfer, 60, 883 

\bibitem[Pi]{Pi} Pilbratt, G.\ et al.\, \aap, in press

\bibitem[Roberge et al.(1991)]{1991ApJS...77..287R} Roberge, W.~G., Jones, 
D., Lepp, S., \& Dalgarno, A.\ 1991, \apjs, 77, 287 

\bibitem[Sch]{Sch} Schmidt, M., Pu{\l}ecka, M., Szczerba, R. et al.\ 2010, in preparation

\bibitem[Sch{\"o}ier 
\& Olofsson(2000)]{2000A&A...359..586S} Sch{\"o}ier, F.~L., \& Olofsson, H.\ 2000, \aap, 359, 586 

\bibitem[Skinner et al.(1999)]{1999MNRAS.302..293S} Skinner, C.~J., 
Justtanont, K., Tielens, A.~G.~G.~M., Betz, A.~L., Boreiko, R.~T., 
\& Baas, F.\ 1999, \mnras, 302, 293
 
\bibitem[Wallerstein 
\& Knapp(1998)]{1998ARA&A..36..369W} Wallerstein, G., \& Knapp, G.~R.\ 1998, \araa, 36, 369 

\bibitem[Willacy(2004)]{2004ApJ...600L..87W} Willacy, K.\ 2004, \apjl, 600, 
L87 

\end{thebibliography}
\end{document}